# Resolving spin, valley, and moiré quasi-angular momentum of interlayer excitons in WSe$_2$/WS$_2$ heterostructures

Chenhao Jin[1]†, Emma C. Regan[1,2]†, Danqing Wang[1,2]†, M. Iqbal Bakti Utama[1,3], Chan-Shan Yang[1,4], Jeffrey Cain[1,5,6], Ying Qin[7], Yuxia Shen[7], Zhiren Zheng[1], Kenji Watanabe[8], Takashi Taniguchi[8], Sefaattin Tongay[7], Alex Zettl[1,5,6], Feng Wang[1,5,6]*

[1] Department of Physics, University of California at Berkeley, Berkeley, California 94720, United States.

[2] Graduate Group in Applied Science and Technology, University of California at Berkeley, Berkeley, California 94720, United States.

[3] Department of Materials Science and Engineering, University of California at Berkeley, Berkeley, California 94720, United States.

[4] Institute of Electro-Optical Science and Technology, and Undergraduate Program of Electro-Optical Engineering, National Taiwan Normal University, Taipei 11677, Taiwan.

[5] Material Science Division, Lawrence Berkeley National Laboratory, Berkeley, California 94720, United States.

[6] Kavli Energy NanoSciences Institute at University of California Berkeley and Lawrence Berkeley National Laboratory, Berkeley, California 94720, United States.

[7] School for Engineering of Matter, Transport and Energy, Arizona State University, Tempe, Arizona 85287, United States.

[8] National Institute for Materials Science, 1-1 Namiki, Tsukuba, 305-0044, Japan.

† These authors contributed equally to this work

* Correspondence to: fengwang76@berkeley.edu

**Moiré superlattices provide a powerful way to engineer properties of electrons and excitons in two-dimensional van der Waals heterostructures[1-11]. The moiré effect can be especially strong for interlayer excitons, where electrons and holes reside in different layers and can be addressed separately. In particular, it was recently proposed that the moiré superlattice potential not only localizes interlayer exciton states at different superlattice positions, but also hosts an emerging moiré quasi-angular momentum (QAM) that periodically switches the optical selection rules for interlayer excitons at different moiré sites[12,13]. Here we report the observation of multiple interlayer exciton states coexisting in a $WSe_2/WS_2$ moiré superlattice and unambiguously determine their spin, valley, and moiré QAM through novel resonant optical pump-probe spectroscopy and photoluminescence excitation spectroscopy. We demonstrate that interlayer excitons localized at different moiré sites can exhibit opposite optical selection rules due to the spatially-varying moiré QAM. Our observation reveals new opportunities to engineer interlayer exciton states and valley physics with moiré superlattices for optoelectronic and valleytronic applications.**

Moiré superlattices between atomically thin materials can dramatically change the properties of electrons and excitons by introducing a new length and energy scale. Artificially stacked moiré superlattices have enabled a variety of intriguing phenomena that are not available in natural systems, such as tunable Mott insulators and unconventional superconductivity [1-11]. Interlayer excitons in van der Waals heterostructures have recently attracted much research interest due to their large binding energy and long lifetime [14,15]. An interlayer exciton is composed of an electron and a hole that are separated in neighboring layers, so its properties can depend strongly on the layer configurations and external fields. For example, it was recently predicted that moiré superlattices, where the interlayer atomic registry changes periodically over space, can host arrays of localized interlayer exciton states with distinct valley selection rules [12,13]. The moiré degree of freedom for interlayer excitons, therefore, offers exciting opportunities towards realizing novel quantum emitter sources and exotic quantum phases like exciton Dirac and Weyl nodes [12,13].

Direct experimental observation of the interlayer moiré excitons with registration-dependent valley selection rules, however, has been challenging. So far, photoluminescence (PL) has been the only probe used to study interlayer excitons due to its sensitivity [16-21], but it has severe limitations because the PL intensity is determined by both the oscillator strength and the excited state lifetime. The sensitive dependence of the excited state lifetime on various sample parameters had led to completely different experimental observations across recent studies and consequently competing theoretical interpretations [16-21]. Optical absorption is a more reliable probe of excitons because it directly measures the transition dipole moment, but its application to interlayer excitons has so far been hindered by the small interlayer exciton oscillator strength [22]. Here we overcome this challenge by probing the absorption spectrum of interlayer excitons in near-zero twist angle WSe$_2$/WS$_2$ heterostructures with background-free techniques: photoluminescence excitation (PLE) and resonant pump-probe spectroscopy. This combination allows for extremely sensitive measurements of the weak absorption features associated with interlayer moiré excitons, and unambiguously determines the nature of the interlayer excitons, including their relative oscillator strengths and the spin-valley configurations of the constituent electrons and holes. We establish that interlayer excitons with the same spin-valley configuration can have opposite circular selection rules, which is attributed to the different moiré quasi-angular momentum (QAM) associated with different interlayer lattice registrations. We further show that an opposite-spin exciton state, which was originally forbidden with no valley selectivity, can gain a well-defined circular helicity in the moiré superlattice.

Figures 1a and 1b show a schematic and an optical microscope image of a representative near-zero twist angle WSe$_2$/WS$_2$ heterostructure (see methods and ref. [23] for device fabrication details). Fig. 1c shows the reflection contrast of the heterostructure in the range of 1.6 to 2.4 eV. Three prominent absorption peaks are observed around the WSe$_2$ A exciton energy of 1.7 eV. This is the characteristic behavior of the intralayer exciton in a moiré superlattice, where the WSe$_2$ A exciton is split into multiple peaks by the strong moiré superlattice potential [23,24]. The moiré superlattice effects on interlayer excitons are expected to be even stronger: the interlayer excitons are predicted to localize at different potential minima within the moiré superlattice, labelled as point A and B in Fig. 1d [12,13]. Consequently, four low energy interlayer moiré exciton states can exist in the K valley, i.e. the same-spin and opposite-spin states centering at A and B points within the moiré superlattice, respectively. These four states are illustrated in Fig. 1d, and their time-reversal pairs create another set of 4 states in the K' valley (not shown).

However, the PL spectrum of the WSe$_2$/WS$_2$ heterostructure shows a single prominent emission peak at 1.43 eV (Fig. 1d), corresponding to only one interlayer exciton state. Although the PL emission spectrum is extremely sensitive, it is not sufficient to probe different interlayer exciton states or the nature of the interlayer excitons: The lowest energy exciton state, which has the longest population lifetime, can dominate the PL emission irrespective of its spin or valley characteristics.

We use helicity-resolved PLE spectroscopy to probe higher energy interlayer moiré exciton states in the WSe$_2$/WS$_2$ heterostructure. We monitor both the $\sigma^+$- and $\sigma^-$-polarized PL emission intensity at 1.43 eV while continuously varying the energy of the excitation photons with $\sigma^+$ helicity. Figure 2a and 2b show the PLE spectra: the energy of $\sigma^+$ excitation light is scanned over the WSe$_2$ intralayer exciton range (1.65 to 1.92 eV, Fig. 2a) and interlayer exciton range (1.45 to 1.55 eV, Fig. 2b). The helicity contrast for intralayer and interlayer excitation ranges are displayed in Fig. 2c and 2d, respectively. The PL emission intensity is proportional to the absorbed photon number and provides a sensitive measurement of the absorption oscillator strength. Strongly-enhanced PL signal is observed when the excitation light is on resonance with the intralayer exciton resonances (Fig. 2a). We also observe well-defined absorption resonances in the interlayer exciton range at 1.46 (green shaded region in Fig. 2b) and 1.51 eV (yellow shaded region). These resonances correspond to two new interlayer moiré exciton states, and their oscillator strengths are more than 100 times smaller than the intralayer exciton transitions. Such weak interlayer exciton transitions are extremely difficult to measure in a direct absorption measurement but are readily observable in the background-free PLE spectroscopy. Furthermore, we observe distinctively different circular helicity behaviors between the intralayer and interlayer excitons: The circular helicity has a large and near-constant value of ~ 0.5 over the whole intralayer exciton range (Fig. 2c), but changes dramatically and can have opposite signs for different interlayer exciton resonances (Fig. 2d, see supplementary).

To resolve the spin and valley properties and to understand the unusual optical selection rules of different interlayer moiré exciton states, we employ resonant pump-probe spectroscopy. As illustrated in Fig. 3a, we resonantly excite an interlayer exciton transition with circularly polarized pump light, and then probe the spin-valley state of the constituent holes in the $WSe_2$ layer by monitoring the induced absorption changes in the $WSe_2$ intralayer exciton transitions. This method takes advantage of the fact that the intralayer exciton optical selection rules have already been well established in previous studies [25-27], and they are independent of the relative registration of the two layers [12,13]. Further, the interlayer exciton oscillator strength can be obtained from the signal magnitude of the resonant interlayer-exciton-pump and intralayer-exciton-probe measurements.

Figure 3b shows the pump-induced circular dichroic signal with probe energy fixed at 1.67 eV (near a $WSe_2$ intralayer exciton feature) and pump energy swept from 1.38 eV to 1.54 eV. Strong pump-probe signals with opposite signs are observed at 1.46 and 1.51 eV (green and yellow shaded regions, respectively). This result reaffirms the PLE observation of two interlayer exciton states at 1.46 and 1.51 eV with different helicity. Interestingly, no clear absorption resonance is observed at the energy of the PL emission peak (purple shaded region around 1.43 eV). This indicates that the interlayer exciton state that dominates the emission process of the system has small absorption oscillator strength. We label it as a "weakly-absorbing" interlayer exciton state, in contrast to the "strongly-absorbing" states at 1.46 and 1.51 eV.

We further resonantly excite the strongly-absorbing interlayer exciton states using $\sigma^+$ pump light and measure the induced circular dichroic spectra in the $WSe_2$ intralayer exciton range (Fig. 3c and 3d). These results can be compared to the pump-probe responses when directly exciting the intralayer exciton at 1.94 eV (Fig. 3e), where the optical selection rule is well established and not affected by the moiré superlattice [12,13]. All three spectral profiles are similar and display prominent resonance features around probe energies of 1.68 eV and 1.73 eV, corresponding to peaks I and II of the intralayer moiré exciton states in $WSe_2$, respectively (Ref. [23]). This is expected because the pump-probe signals originate from the response of intralayer excitons to valley-polarized holes in $WSe_2$. The sign of the signals, however, are different for the two interlayer exciton states. Pumping at 1.51eV gives a signal of the same sign as pumping the intralayer exciton. This interlayer exciton state therefore has the same optical selection rule as the intralayer exciton, i.e. $\sigma^+$ light selectively creates holes in the K valley of $WSe_2$ (inset of Fig. 3d and Fig. 3e). In contrast, pumping at 1.46 eV gives the opposite sign, indicating that $\sigma^+$ light selectively create holes in the K' valley (inset of Fig. 3c). In other words, the 1.51 eV and 1.46 eV interlayer exciton states in the K valley will couple more efficiently to $\sigma^+$ and $\sigma^-$ light, and therefore have a total quasi-angular momentum (QAM) of +1 and -1, respectively (see supplementary).

Based on the spin-valley state of the hole in the $WSe_2$ layer and the interlayer exciton oscillator strength, we can infer the spin-valley configuration of the electron within the interlayer exciton and determine the emerging moiré quasi-angular momentum, as shown in Table 1. The electron and hole must be in the same valley and of the same spin for the "bright" 1.46 and 1.51 eV interlayer exciton states to exhibit relatively strong oscillator strengths. Therefore, we can assign a spin contribution of 0 and valley contribution of +1 from the constituent electron and hole to the QAM of K-valley interlayer exciton (see supplementary). The rest of the contribution to the QAM arises from the local interlayer atomic registry in the moiré superlattice. Consequently, we determine a moiré QAM of -2 for the 1.46 eV interlayer exciton and zero for the 1.51 eV interlayer exciton. They correspond to the second (moiré position B) and fourth state (moiré position A) illustrated in Fig. 1d, respectively.

The lowest energy interlayer exciton state at 1.43 eV has weak oscillator strength, and we attribute it to an interlayer exciton with opposite electron and hole spin based on the electronic bands of the heterostructure (Fig. 1d). This arises naturally because the spin-orbital coupling has opposite sign in the conduction and valence bands for W-based transition metal dichalcogenide materials [28,29]. This opposite spin state is not completely dark, since spin is no longer a good quantum number in systems with strong spin-orbital coupling [30-32]. The observed energy separation between the opposite- and same-spin interlayer excitons (state 1 and 2) is ~30 meV, which is consistent with the known conduction band spin splitting in $WS_2$ layers [28,29]. Previously, an intralayer opposite-spin exciton has been observed in the PL emission of monolayer $WSe_2$ without a magnetic field [33,34]; however, it shows no circular valley selection, as its QAM only comes from spin-valley contribution and is 0 (Ref. [30,31]). In contrast, emission from the 1.43 eV interlayer exciton state shows a large $\sigma^+$ circular helicity when we excite intralayer excitons with $\sigma^+$-polarized light (Fig. 2b). It indicates that the lowest energy K-valley interlayer exciton has a total QAM of +1 (state 1 in Fig. 1d). This again highlights the important role of moiré superlattice, which introduces an additional QAM of +1 (or equivalently, -2, due to the three-fold rotation symmetry in the system) and a well-defined circular selection rule to the lowest-energy opposite-spin state. The other opposite-spin state, i.e. state 3, is not observed in our study, presumably due to its very small oscillator strength.

Combining the ability to engineer spin-valley selection rules using the novel moiré degree of freedom, the coexistence of multiple moiré states with distinctively different optical properties, and the tunability from layer configuration and external fields, interlayer excitons in $WSe_2/WS_2$ moiré superlattice provide a versatile platform for exciting new exciton and valley physics.

**Methods:**

Heterostructure preparation

Monolayer WSe$_2$, monolayer WS$_2$, few-layer graphene, and thin hBN flakes were exfoliated onto silicon substrates with a 90 nm silicon oxide layer. Polarization-dependent second harmonic generation (SHG) was used to determine the relative angles between the WSe$_2$ and WS$_2$ crystalline axes (see supplementary text and Ref. [23]). The WSe$_2$/WS$_2$ heterostructures were prepared using a dry transfer method with a polyethylene terephthalate (PET) stamp[35]. A PET stamp was used to pick up the top hBN flake, the WS$_2$ monolayer, the WSe$_2$ monolayer, a few-layer graphene electrode, the bottom hBN, and the few-layer graphene back gate at 60 °C. The angle of the PET stamp was adjusted manually between picking up the WS$_2$ and WSe$_2$ flakes to ensure that the angle between the crystal axes was near zero. The PET stamp was then stamped onto a clean Si substrate with 90 nm SiO$_2$ at 130 °C, and the PET was dissolved in dichloromethane for 12 hours at room temperature. Contacts (~75 nm gold with ~5 nm chromium adhesion layer) to the few-layer graphene flakes were made using electron-beam lithography and electron-beam evaporation.

Photoluminescence excitation spectroscopy

PLE measurements were performed using a pulsed supercontinuum laser source (Fianium FemtoPower 1060), which was spectrally filtered by a grating and appropriate filters to provide a tunable excitation light with 0.3 nm linewidth. The $\sigma^+$ polarization state of the excitation beam was set with a linear polarizer and quarter wave plate. The excitation light was focused at the sample with ~ 3 um beam size. $\sigma^+$- and $\sigma^-$-polarized PL was collected at each excitation energy and analyzed with a monochromator and a liquid-nitrogen-cooled charge coupled device (CCD).

The PL spectra were normalized to excitation power and integration time. We verified that the PL intensity scales linearly with excitation power for all energies studied. The integrated PL peak intensity was then plotted as a function of the excitation energy.

Resonant pump-probe spectroscopy

Femtosecond pulses at 1026 nm (repetition rate of 150 kHz and duration of ~300 fs) were generated by a regenerative amplifier that is seeded by a mode-locked oscillator (Light Conversion PHAROS). The femtosecond pulses were split into two paths: one was used to pump an optical parametric amplifier (Light Conversion TOPAS). Its signal output was then used as the pump light in our measurement. The second portion of the 1026 nm beam was focused onto a sapphire crystal to generate broadband white light, which was then spectrally filtered by appropriate bandpass filters (each with 10 nm FWHM) to serve as the probe light. The pump and probe polarizations were set with linear polarizers and a shared quarter wave plate. The pump-probe delay time was controlled with a motorized delay stage. The pump and probe beams were focused at the sample with diameters of 50 um and 25 um, respectively. The reflected probe light was detected by a photomultiplier tube after wavelength selection using a monochromator (passing bandwidth of 5 nm). The pump-probe signal was analyzed using a lock-in amplifier at a ~ 2.5 kHz modulation frequency.

**Acknowledgements**:

This work was supported primarily by the Director, Office of Science, Office of Basic Energy Sciences, Materials Sciences and Engineering Division of the U.S. Department of Energy under contract no. DE-AC02-05-CH11231 (van der Waals heterostructures program, KCWF16). PLE spectroscopy of the heterostructure is supported by the US Army Research Office under MURI award W911NF-17-1-0312. Growth of hexagonal boron nitride crystals was supported by the Elemental Strategy Initiative conducted by the MEXT, Japan and JSPS KAKENHI Grant Numbers JP15K21722. S.T. acknowledges the support from NSF DMR 1552220 NSF CAREER award for the growth of $WS_2$ and $WSe_2$ crystals. E.C.R acknowledges support from the Department of Defense (DoD) through the National Defense Science & Engineering Graduate Fellowship (NDSEG) Program. C.S.Y. acknowledges support from the grant 107-2112-M-003-014-MY3 of the Ministry of Science and Technology.

**Author contributions:**

C.J., E.C.R., and D.W. contributed equally to this work. F.W. and C.J. conceived the research. C.J., E.C.R. and C.S.Y. built the optical setup. C.J., E.C.R. and D.W. carried out optical measurements. C.J., F.W. and E.C.R. performed theoretical analysis. E.C.R., D.W., M.I.B.U. and Z.Z. fabricated van der Waals heterostructures. Y.Q., S.Y., S.T., J.C., and A.Z. grew $WSe_2$ and $WS_2$ crystals. K.W. and T.T. grew hBN crystals. All authors discussed the results and wrote the manuscript.

**Competing interests:**

The authors declare no competing interests.

**Data and materials availability:** The data that support the findings of this study are available from the corresponding author upon reasonable request.

Figures:

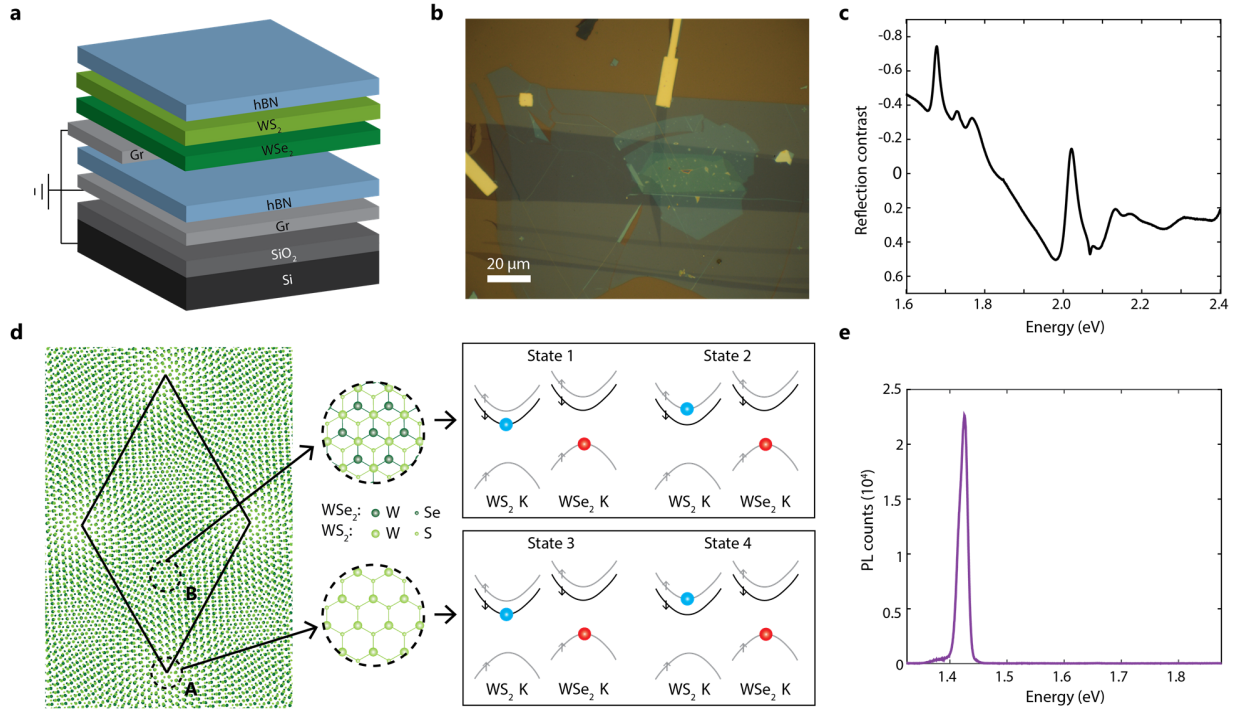

**Fig. 1 | Interlayer moiré excitons in near-zero twist angle WSe$_2$/WS$_2$ heterostructure.** (**a** and **b**) Side-view illustration (a) and optical microscope image (b) of a representative near-zero twist angle heterostructure. (**c**) Reflection contrast of the heterostructure shows three prominent peaks in the WSe$_2$ A exciton range near 1.7 eV. This is a characteristic absorption signature of the intralayer moiré exciton. (**d**) Illustration of the moiré superlattice in real space (left) with the moiré supercell outlined in a black diamond. Interlayer excitons can be trapped at two different local minima of the moiré potential, labeled as A and B points. This moiré degree of freedom, combined with different spin configurations, gives rise to 4 interlayer exciton states in the K valley (state 1 to 4 in the right box). (**e**) Photoluminescence (PL) spectrum of the heterostructure shows only one prominent peak at 1.43 eV because PL measurements are only sensitive to the emission state and cannot probe higher energy states.

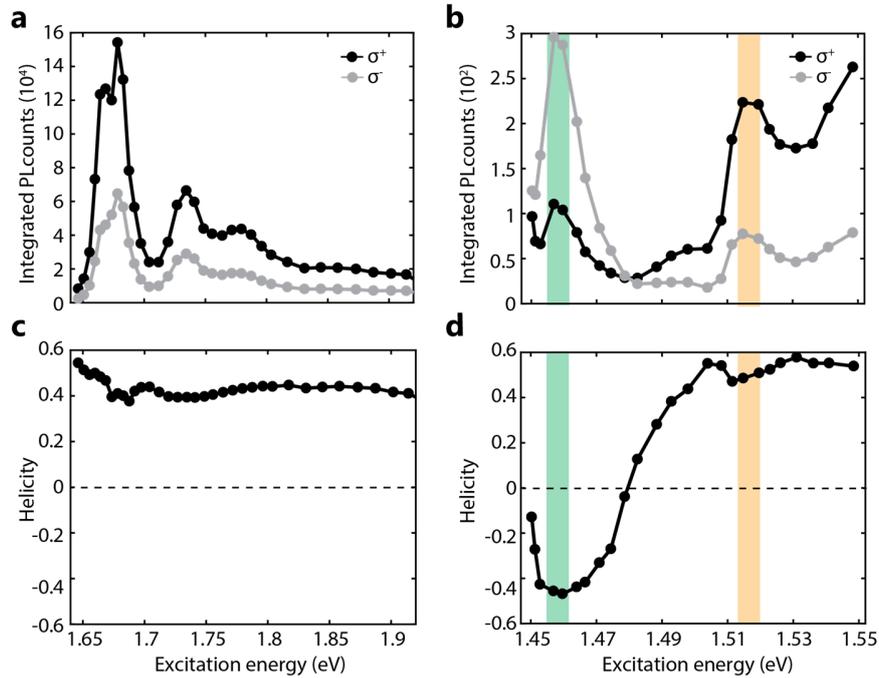

**Fig. 2 | Interlayer moiré excitons probed by helicity-resolved photoluminescence excitation (PLE) spectroscopy.** (**a** and **b**) PLE spectra of a representative device measured by monitoring the $\sigma^+$ (black) and $\sigma^-$ (grey) emission intensity of the 1.43 eV emission peak. The energy of $\sigma^+$ excitation light is scanned over intralayer exciton (a) and interlayer exciton (b) range. The emission intensity is strongly enhanced when excitation light is in resonance with all intralayer exciton resonances (a) and two additional resonance peaks in the interlayer exciton range (b), suggesting the existence of two new interlayer moiré exciton states at 1.46 eV (green shaded area) and 1.51 eV (yellow shaded area), respectively. (**c**) PL circular helicity shows a near-constant positive value of ~ 0.5 with intralayer exciton excitation. (**d**) PL emission shows circular helicity of ~ -0.5 and ~ 0.5 when exciting the 1.46 and 1.51 eV interlayer exciton states, revealing their unusual optical selection rules. All measurements are done at 10 Kelvin.

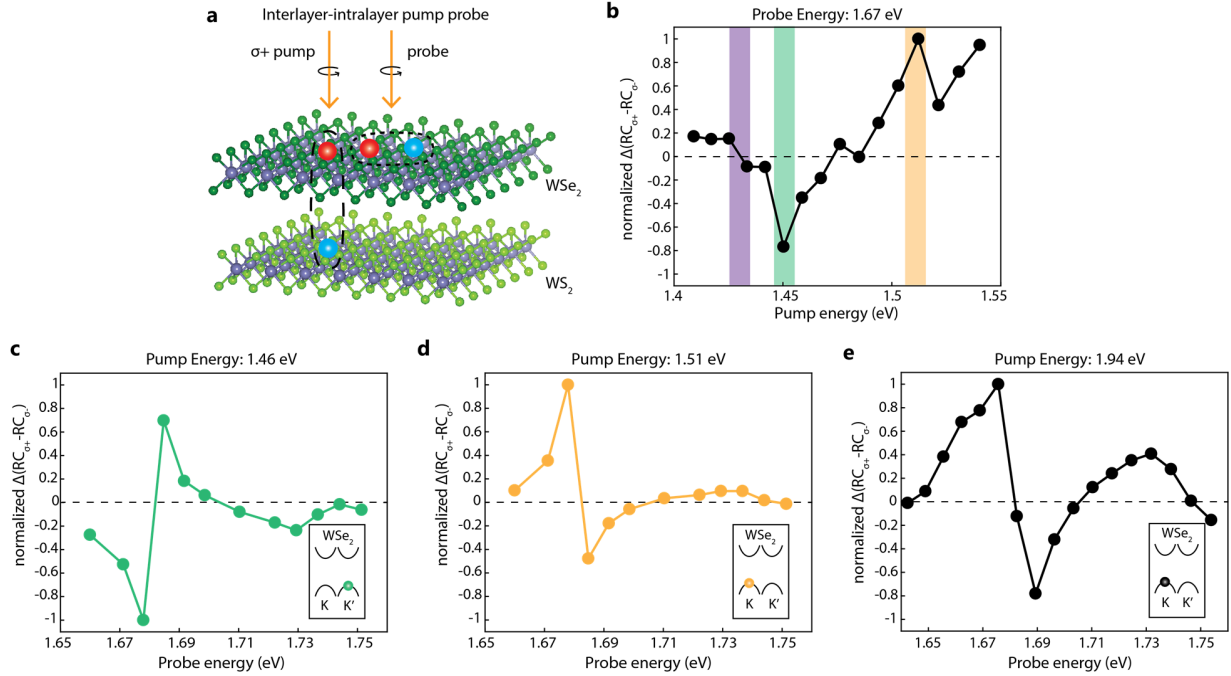

**Fig. 3 | Interlayer moiré excitons probed by resonant pump-probe spectroscopy.** (**a**) Illustration of the pump-probe experimental design. σ⁺ pump light in resonance with an interlayer exciton state directly creates valley-polarized holes in WSe$_2$, which can then be probed by the induced circular dichroic signal at the WSe$_2$ intralayer exciton energy. (**b**) Pump-induced circular dichroic signal with probe energy fixed at 1.67 eV and pump energy swept from 1.38 eV to 1.54 eV. The two prominent resonances with opposite sign at 1.46 (green shaded region) and 1.51 eV (yellow shaded region) reaffirm the two interlayer moiré exciton states observed in PLE measurement. The 1.43 eV state, on the other hand, shows no clear resonance (purple shaded region), indicating its weakly-absorbing nature. (**c** to **e**) Pump-induced circular dichroic spectra with pump energy in resonance with the 1.46 eV state (c) and 1.51 eV state (d), and in the intralayer exciton range at 1.94 eV (e). The circular dichroic spectra when pumping the 1.51 eV interlayer exciton state gives the same sign as when pumping the intralayer exciton; therefore in both cases σ⁺ pump create holes in the K valley of WSe$_2$ (insets of (d) and (e)). In contrast, an opposite sign is observed when pumping the 1.46 eV state, indicating that σ⁺ light selectively create holes in the K' valley (insets of (c)). All measurements are done at 10 Kelvin.

| State | Energy (eV) | Oscillator Strength | Total QAM | Hole | Electron | Spin QAM | Valley QAM | **Moiré QAM** | **Moiré Position** |
|---|---|---|---|---|---|---|---|---|---|
| 1 | 1.43 | Weak | +1 = -2 | K ↑ | K ↓ | -1 | +1 | **-2** | **B** |
| 2 | 1.46 | Strong | -1 | K ↑ | K ↑ | 0 | +1 | **-2** | **B** |
| 3 |  | Not observed | 0 | K ↑ | K ↓ | -1 | +1 | **0** | **A** |
| 4 | 1.51 | Strong | +1 | K ↑ | K ↑ | 0 | +1 | **0** | **A** |
| 5 | 1.43 | Weak | -1 = +2 | K' ↓ | K' ↑ | 1 | -1 | **+2** | **B** |
| 6 | 1.46 | Strong | +1 | K' ↓ | K' ↓ | 0 | -1 | **+2** | **B** |
| 7 |  | Not observed | 0 | K' ↓ | K' ↑ | 1 | -1 | **0** | **A** |
| 8 | 1.51 | Strong | -1 | K' ↓ | K' ↓ | 0 | -1 | **0** | **A** |

**Table 1:** The nature of different interlayer moiré exciton states and the spin, valley and moiré contributions to their optical selection rules

**Supplementary Information for**

**Resolving spin, valley, and moiré quasi-angular momentum of interlayer excitons in $WSe_2/WS_2$ heterostructures**

Chenhao Jin†, Emma C. Regan†, Danqing Wang†, M. Iqbal Bakti Utama, Chan-Shan Yang, Jeffrey Cain, Ying Qin, Yuxia Shen, Zhiren Zheng, Kenji Watanabe, Takashi Taniguchi, Sefaattin Tongay, Alex Zettl, Feng Wang*

† These authors contributed equally to this work

* Correspondence to: fengwang76@berkeley.edu

**S1. Determination of the rotational alignment between the $WSe_2$ and $WS_2$ layers**

**S2. Decay dynamics in the resonant pump-probe measurements**

**S3. Definition of the exciton quasi-angular momentum**

**S4. Optical selection rules of interlayer excitons**

## S1. Determination of the rotational alignment between the WSe$_2$ and WS$_2$ layers

The crystal orientation of the WS$_2$ and WSe$_2$ monolayer flakes are determined by angle-resolved SHG measurements before the heterostructure fabrication. During the transfer process, the relative angle between the WS$_2$ and WSe$_2$ layer is controlled to be near-zero within 0.5-degree accuracy. Afterwards, the alignment between the WSe$_2$ and WS$_2$ layers is confirmed by directly measuring the SHG signal from the heterostructure, as shown in Fig. S1. The heterostructure SHG intensity is about four times larger than that of the monolayer WSe$_2$ and WS$_2$ regions measured at the same experimental condition. This indicates that the WSe$_2$ and WS$_2$ layers have near-zero twist angle (AA stacking) instead of near-60 twist angle (AB stacking).

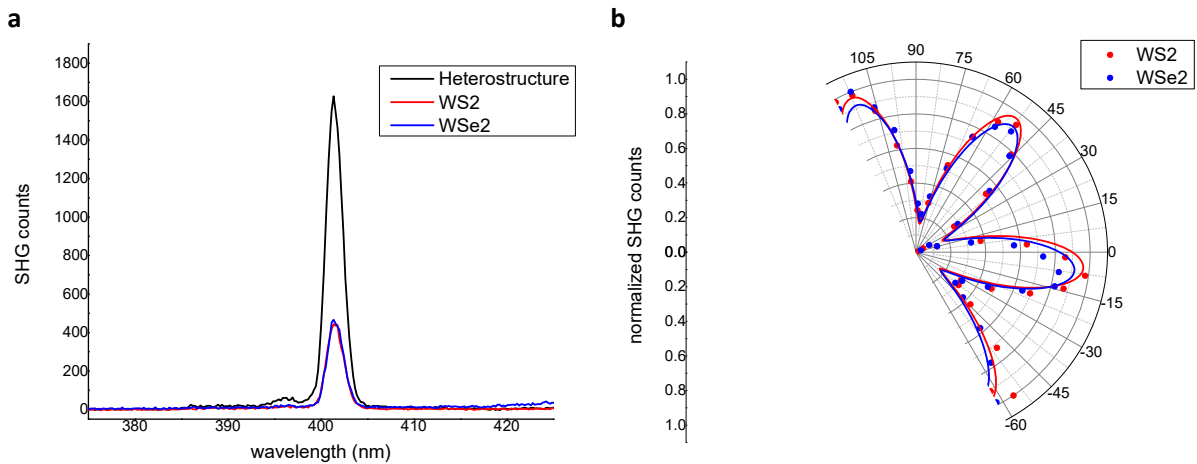

**Fig. S1.** (**a**) SHG signal of the device measured on the heterostructure (black), WS$_2$ alone (red), and WSe$_2$ alone (blue) regions with the same experimental configuration. (**b**) Polarization-dependent SHG signal measured on monolayer WS$_2$ (red circles) and WSe$_2$ (blue circles) regions and corresponding fittings (red and blue curves).

## S2. Decay dynamics in the resonant pump-probe measurements

Figure S2 shows a representative transient circular dichroic signal with 1.51 eV pump (resonant with interlayer exciton state 4) and 1.67 eV probe energy (near the WSe$_2$ intralayer exciton). The signal shows an immediate rise at time-zero, and remains nearly a constant afterwards with decay lifetime much longer than the experimental delay range of 300 ps. Such dynamics can be understood as the following: The circular dichroic signal originates from valley-polarized holes in WSe$_2$ (Ref. [36]), which are directly created by the pump light. In addition, valley-polarized holes can have microseconds-long lifetime in the heterostructure, and therefore show negligible decay in the experimental delay range[36]. The sharp peak at time zero is due to the coherent optical Stark effect[37]. Because the pump-probe signal is nearly constant from 10 to 300 ps, we can choose any pump-probe delay within this range to evaluate the density of valley-polarized holes in WSe$_2$. For Fig. 3 in the text, all data points are obtained at pump probe delay of 200 ps.

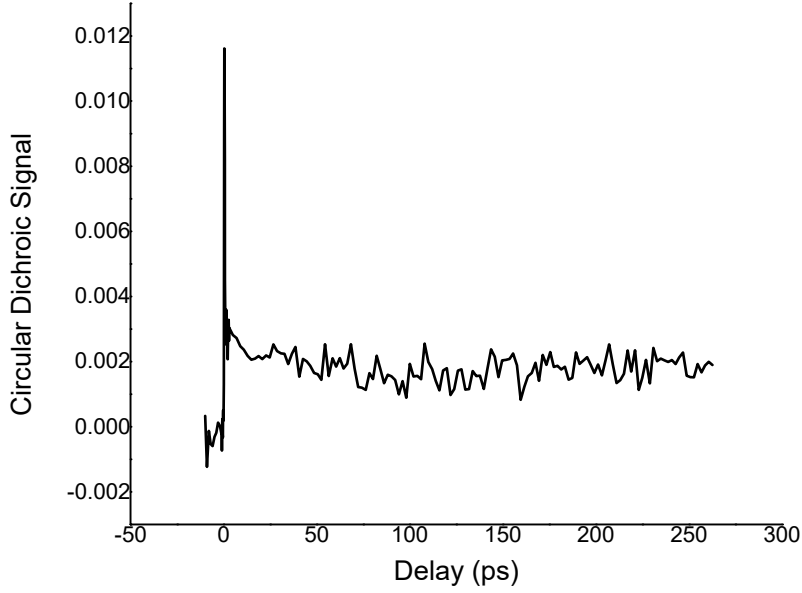

**Fig. S2.** Representative dynamics of resonant pump probe signal. The signal behaves like a step function, which rises immediately at the beginning and remains a constant afterwards. The sharp spike at time zero is due to the coherent optical Stark effect.

### S3. Definition of the exciton quasi-angular momentum

The WSe$_2$/WS$_2$ bilayer has three-fold rotational ($\hat{C}_3$) symmetry in real space. As a result, excitons at high-symmetry points in momentum space (for example, in K or K' valleys and with zero total momentum) must be eigenfunctions of the $\hat{C}_3$ operation:

$$\hat{C}_3 \Psi = e^{i\frac{2\pi}{3}n}\Psi. \quad (S1)$$

Here, $\Psi$ is the exciton wavefunction, and $n$ is the exciton quasi-angular momentum (QAM) which describes the phase change of the exciton wavefunction after a 120-degree rotation. For bosons, three continuous $\hat{C}_3$ operations should bring $\Psi$ back, so $n$ must be an integer. In addition, $n$ and $(n + 3m)$ describes identical situation ($m$ is an arbitrary integer). Therefore, there are only three independent QAM values: $n = 0, 1$ and $2$. For convenience, we have also used $n = -1$ (equivalent to +2) and $n = -2$ (equivalent to +1) in the text.

Following similar definition, σ$^+$ and σ$^-$ light carry a QAM of +1 and -1, respectively. Because the ground state always has QAM of 0 and the QAM must be conserved during the optical transition, the σ$^+$ (σ$^-$) light can only couple to exciton state with QAM of +1 (-1).

## S4. Optical selection rules of interlayer excitons

The optical selection rule of an exciton state is directly determined by its total QAM. Different contributions to the interlayer exciton QAM can be obtained from the definition in Eq. S1. The exciton wavefunction can be expanded as

$$\Psi(\mathbf{r}_e, \mathbf{r}_h) = \psi(\mathbf{R})\phi(\mathbf{r}).$$

Here $\mathbf{r}_e$ and $\mathbf{r}_h$ are the coordinates for electron and hole within the exciton; $\mathbf{R}$ and $\mathbf{r}$ are the center-of-mass coordinate and relative coordinate of the quantum two-body system, respectively. $\phi(\mathbf{r})$ describes the relative motion between the electron and hole within the exciton, which gives rise to a series of hydrogen-like levels such as 1s and 2p states. Because the exciton binding energy is large in two-dimensional transition metal dichalcogenide (TMDC) heterostructures[38,39], here we only consider the 1s state, which has zero total angular momentum from $\phi(\mathbf{r})$. The QAM of the 1s exciton is therefore solely coming from $\psi(\mathbf{R})$, the center-of-mass motion part of the exciton wavefunction.

In this case, the QAM of the exciton can be intuitively understood from the QAM of the electron and hole states at the respective band edges. For intralayer excitons in monolayer TMDC materials, it is well-established that same-spin and opposite-spin excitons in the K (K') valley have a total QAM of +1 (-1) and 0, respectively, from the spin-valley contribution[31,32]. Therefore the same-spin exciton in the K (K') valley only couples to σ+ (σ-) light, while the opposite-spin exciton does not have valley absorption selection rule. These behaviours are not affected by adding another layer nearby, since the electron and hole are in the same layer and always experience the same local environment.

On the other hand, interlayer excitons, where the electron and hole reside in separate layers, can have different local symmetry depending on the atomic registry between the two layers. At point A of the moiré superlattice, the two layers completely overlap (i.e. the metal and chalcogen atoms of one layer are exactly on top of the corresponding atoms of the other layer), so the local environment for the electron and hole is again the same, generating no additional QAM. While at point B, an additional moiré contribution of -2 (+2) is introduced to the QAM for interlayer excitons in the K (K') valley[12,13]. Their optical selection rules can then be obtained by summing up the spin-valley contribution and moiré contribution to the total QAM. For positions away from the high symmetry points in the moiré pattern, the optical selection rule becomes non-perfect as the $\hat{C}_3$ symmetry is locally broken. Therefore, an interlayer exciton will have well-defined circular selection rule only if its wavefunction is highly localized around the high symmetry points within the moiré superlattice[12,13].

One can estimate the length scale of the localization by inspecting how well-defined the optical selection rule is in PLE results. The near-constant circular helicity of ~ 0.5 over the whole intralayer exciton range (Fig. 2c) is consistent with previous studies showing that the absorption valley selection rule in monolayer $WSe_2$ is near-perfect over a large excitation energy range and that σ+ light almost solely creates holes in the K valley[25]. The amplitude of the PL circular helicity is also quite large for excitation in the interlayer exciton range, reaching ~ 0.5 and ~ -0.5, for the same-spin states. When compared to the case of intralayer exciton range, these numbers suggest that holes are almost solely created in the K or K' valley when directly exciting the two interlayer exciton states with σ+ light. The well-defined optical selection rules of interlayer moiré exciton states require that their center-of-mass wavefunctions be concentrated in a length scale that is much smaller than the moiré periodicity, forming multiple well-separated exciton lattices with distinctively different spin and valley selection rules.